\documentstyle[twocolumn,prl,aps,epsfig]{revtex}
\newcommand{\bfr}{{\bf r}}

\def\pra#1#2#3{Phys.~Rev.~A~{\bf #1},\ #2\ (#3)}
\def\prl#1#2#3{Phys.~Rev.~Lett.~{\bf #1},\ #2\ (#3)}
\def\sci#1#2#3{Science~{\bf #1},\ #2\ (#3)}
\def\nat#1#2#3{Nature~{\bf #1},\ #2\ (#3)}
\begin{document}
\flushbottom \draft
\title{Controlling two-species Mott-insulator phases in an optical lattice to form
an array of dipolar molecules}
\author{M. G. Moore \cite{emailm}  and H. Sadeghpour\cite{emails}}
\address{ITAMP, Harvard-Smithsonian Center for Astrophysics, Cambridge, MA 02138 \\ (\today)
\\ \medskip}\author{\small\parbox{14.2cm}{\small\hspace*{3mm}
We consider the transfer of a two-species Bose-Einstein condensate into an optical lattice
with a density such that that a Mott-insulator state with one atom per species per lattice
site is obtained in the deep lattice regime. Depending on collision parameters the result
could be either a `mixed' or a `separated' Mott-insulator phase. Such a `mixed' two-species
insulator could then be photo-associated into an array of dipolar molecules suitable for
quantum computation or the formation of a dipolar molecular condensate. For the case of a
$^{87}$Rb-$^{41}$K two-species BEC, however, the large inter-species scattering length makes
obtaining the desired `mixed' Mott insulator phase difficult. To overcome this difficulty we
investigate the effect of varying the lattice frequency on the mean-field interaction and
find a favorable parameter regime under which a lattice of dipolar molecules could be
generated.
\\
\\[3pt]PACS numbers: 03.75.Fi, 03.75.Be, 03.75.-b }}
\maketitle

\maketitle \narrowtext


In a recent experiment \cite{GreManEss02} a quantum phase transition from a superfluid (SF)
to a Mott insulator (MI) state was observed by varying the depth of a three-dimensional
optical lattice \cite{JakBruCir98} superimposed onto a trapped Bose-Einstein condensate
(BEC) of $^{87}$Rb atoms. This experiment highlights a growing trend in ultracold atomic
physics whereby condensates are no longer the direct object of study, but serve mainly as
well-controlled initial state for the preparation of more exotic highly-correlated many-body
states. In addition to providing new insight into fundamental phenomena of condensed-matter
physics, MI states are expected to have important applications in Heisenberg-limited
atom-interferometry \cite{orzel} and quantum computing \cite{BreCavJes99,JakBriCir99}.

Another interesting application of a Mott insulator was recently proposed by Jaksch and
coworkers: employing a Mott insulator state as an intermediate stage in the generation of a
BEC of molecules \cite{JakVenCir02}.  Molecules would be formed via stimulated Raman
photoassociation of a Mott insulator with two atoms per lattice site, resulting in a
molecular MI with unit filling factor which could then be `melted' to form a molecular BEC.
The primary advantage of this approach is that during photoassociation the resulting
molecules would be completely isolated from collisions with other atoms and/or molecules.
This is a distinct advantage over photoassociation in the superfluid phase
\cite{heinzen,wynar}, as photoassociation strongly favors the creation of vibrationally
excited molecules- due to favorable Franck-Condon transition from the continuum-
which tend to undergo inelastic collisions in the presence of other
atoms and/or molecules. The resulting release of vibrational quanta leads to loss of
both particles from the trap. By isolating each vibrationally excited molecule in a lattice site, it should be
possible to stimulate the molecules into the ground rotational-vibrational state before
`melting' the Mott insulator, thus eliminating this collisional loss mechanism.

\begin{figure}
\begin{center}
\epsfig{file=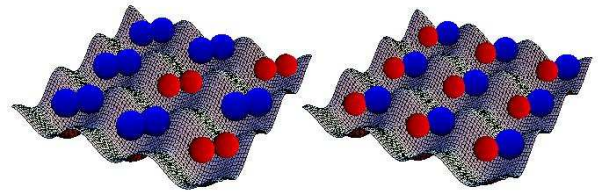,width=8cm}
\end{center}
\caption{Separated- (left) and mixed- (right) two-species Mott insulator phases in an
optical lattice. The separated-phase MI is randomly distributed. } \label{fig1}
\end{figure}

In this Letter, we extend this idea to the formation of a lattice of {\it two-species Mott
insulator phase} with precisely one atom per species per lattice site. This scheme would
require  a two-species Bose-Einstein condensate as a starting point, as has been recently
demonstrated with $^{87}$R and $^{41}$K \cite{ModFerRoa01,ModModRib02}. The two-species
condensate would then be adiabatically loaded into a three-dimensional optical lattice,
resulting in the formation of either a `mixed' or `separated' MI state, corresponding to
each lattice site containing different or identical species, respectively, see Fig.
\ref{fig1}. The formation of a mixed phase turns out to be difficult for the case of a
$^{87}$Rb-$^{41}$K system, due to the large positive inter-species scattering length which
favors the separated phase. We therefore concentrate on overcoming this difficulty by
varying the lattice frequency, which exploits the difference in atomic resonance frequencies
to control the relative densities of the two species.

Preparing a Mott insulator state of heteronuclear molecules by this approach would be an
elegant method to obtain a lattice of dipolar molecules for use as a quantum computer
\cite{Dem02}. In addition a rich variety of quantum phases
\noindent are predicted for the ground state of a lattice of dipolar bosons , including a
supersolid phase \cite{GorSanLew02}. The lattice of dipolar molecules could then be `melted'
by adiabatically lowering the lattice potential, resulting in a condensate of dipolar
bosons. Here the long-range dipole-dipole interactions should introduce interesting
correlations effects in both the ground state and collective excitations of the BEC
\cite{GorRzaPfa00,SanShlZol00,YiYou00,YiYou01,MarMacSuo01}.

Others have considered the related problem of superfluid-insulator transitions for
spinor (multicomponent) BECs \cite{DemZho02} in which the added freedom of superpositions of
hyperfine states is included. One crucial difference between a two-species system and a
spinor system is the fact that each atomic species will see a different optical potential as
a result of having different dipole moments, resonance frequencies, and atomic masses. This
additional freedom should result in the physics of a two-species system being significantly
richer than that of a single-species spinor system. For example, due to the potential for
different tunneling rates between neighboring lattice sites and/or collision rates for each
species, the superfluid-insulator transition will most likely occur at a different lattice
intensity for each species, resulting in a regime where a MI of one species is embedded in a
SF of another species. Thus the phases of a dual-species lattice can be divided into three
categories: a dual SF, a hybrid SF-MI phase, and a dual MI phase. At present we will focus
primarily on the properties of the dual MI phase, as it is the only phase with the potential
for collisionless photo-association into dipolar molecules.

The starting point for our theoretical consideration is a two-species atomic system in the
presence of an optical lattice potential. The atomic density assumed to be such that at the
center of the trap there is one atom per species per cubic wavelength. The Bose-Hubbard
Hamiltonian \cite{JakBruCir98} for such a system is then given by
\begin{eqnarray}
\label{BH}
    \hat{H}&=&\frac{\hbar}{2}\sum_j\left[g_{11}\ \hat{n}_{1j}(\hat{n}_{1j}-1)
    +g_{22}\ \hat{n}_{2j}(\hat{n}_{2j}-1)\right. \nonumber\\
    &+&\left. 2g_{12}\ \hat{n}_{1j}\hat{n}_{2j}\right]
    +\hbar\sum_{i,j}\left[\beta_{1ij}\ \hat{c}^\dag_{1i}\hat{c}_{1j}
    +\beta_{2ij}\ \hat{c}^\dag_{2i}\hat{c}_{2j}\right],
\end{eqnarray}
where $\hat{c}_{ij}$ is the bosonic annihilation operator for species $i$ at lattice site
$j$, $\hat{n}_{ij}=\hat{c}^\dag_{ij}\hat{c}_{ij}$ is the corresponding number operator,
$\beta_{ijk}$ is the rate of tunneling of species $i$ from lattice site $j$ to site $k$, and
$g_{ij}$ is the collision rate for collisions between species $i$ and $j$.

In order to compute the various interaction and tunneling rates we must first consider the
dependence of the optical lattice potential on the intrinsic properties of each atomic
species. The optical potential seen by species $i$ can be expressed as
\begin{equation}
\label{Vir}
    V_i(\bfr)=\frac{d^2_i}{(\omega_i-\omega_L)}\ U(\bfr),
\end{equation}
where $d_i$ and $\omega_i$ are the dipole moment and resonance frequency of species $i$,
respectively, $\omega_L$ is the laser frequency of the optical field, and
$U(\bfr)=-|E(\bfr)|^2/\hbar$ is the species-independent intensity of the optical field.
Assuming that the sign of detuning is the same for both species, in which case the potential
minima will coincide, we can expand around the potential minima according to
$U(\bfr)=u_0+\frac{1}{2}\sum_\mu u_\mu r^2_\mu$, giving
\begin{equation}
\label{Vir2}
    V_i(\bfr)=\frac{1}{2}\frac{d^2_i}{(\omega_i-\omega_L)}\sum_\mu u_\mu r^2_\mu,
\end{equation}
where we have dropped the constant term consistent with a transformation to a rotating frame
which leaves the Hamiltonian (\ref{BH}) unchanged.

Following the expansion in (\ref{Vir2}), we make a Gaussian
approximation for the lowest energy Wannier modes of a lattice site for species $i$ as
\begin{equation}
\label{psii}
    \phi_{i}(\bfr)=\pi^{-3/4}\prod_\mu\lambda^{-1/2}_{i\mu}
    e^{-\frac{1}{2}\left(r_\mu/\lambda_{i\mu}\right)^2},
\end{equation}
where we have introduced the harmonic oscillator length for species $i$ along the direction
$\hat{r}_\mu$
\begin{equation}
\label{lambdaimu}
    \lambda_{i\mu}=\left[\frac{\hbar^2|\omega_i-\omega_L|}{m_id^2_iu_\mu}\right]^{1/4},
\end{equation}
$m_i$ being the atomic mass of species $i$. With these wavefunctions, we can compute the
collision coefficients via
\begin{eqnarray}
\label{gij}
    g_{ij}&=&\frac{2\pi\hbar\ a_{ij}}{\mu_{ij}}
    \int d\bfr\ |\phi_i(\bfr)|^2|\phi_j(\bfr)|^2\nonumber\\
    &=&\frac{2\hbar\  a_{ij}}{\sqrt{\pi}\mu_{ij}}\prod_\mu
    \frac{1}{\sqrt{\lambda^2_{i\mu}+\lambda^2_{j\mu}}}
\end{eqnarray}
where $a_{ij}$ is the scattering length for collisions between species $i$ and $j$, and
$\mu_{ij}=m_im_j/(m_i+m_j)$ is the effective mass.

\begin{figure}
\begin{center}
\epsfig{file=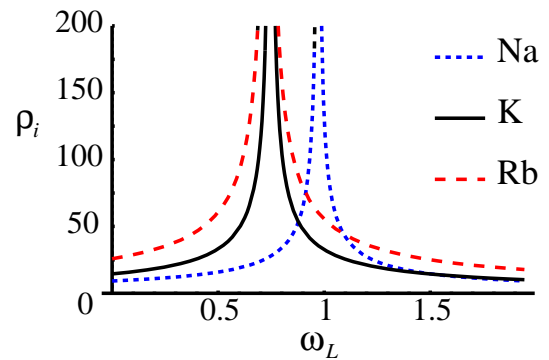,width=7cm}
\end{center}
\caption{Relative density parameter in an optical lattice for $^{23}$Na, $^{41}$K, and
$^{87}$Rb as a function of the lattice frequency $\omega_L$. The relative density parameter
is in units of $\left[m_p e^2a_0^2/cR_\infty\right]^{3/4}$ and the laser frequency is in
units of $cR_\infty$.} \label{fig2}
\end{figure}

Because the harmonic oscillator length (\ref{lambdaimu}) for a given atomic species tends to
zero when approaching the atomic resonance frequency, tuning the lattice frequency close to
the resonance frequency of one atomic species can act as a very strong `handle' with which
to tune the relative atomic densities, and hence the collision parameters of the system. The
local density per atom, defined as $1/V_i=\prod_\mu\lambda_{i\mu}^{-1}$, is given by
$1/V_i=\rho_i\hbar^{-3/2}\prod_\mu u_\mu^{1/2}$, where we have introduced the relative
density parameter
\begin{equation}
\label{rhoi}
    \rho_i=\left[\frac{m_id_i}{|\omega_L-\omega_i|}\right]^{3/4}.
\end{equation}
From the equation for $1/V_i$, we see that the relative densities of the different atomic
species will be independent of the lattice parameters $u_\mu$, provided only that the wells
are sufficiently deep that the Gaussian approximation holds for the lowest Wannier state-
generally required for the lattice depth to be large compared to the atomic recoil energy.
In Figure \ref{fig2}, we plot the dimensionless density parameter $\rho_i$ as a function of
$\omega_L$ for atomic species currently used in BEC experiments: $^{23}$Na, $^{41}$K, and
$^{87}$Rb. From this figure, we note that a wide range of relative densities can be obtained
by varying the lattice frequency across the atomic resonance frequencies. The atomic
parameters in this calculation are: $m_i=\{22.99,40.96,86.91\}\times m_p$,
$\omega_i=\{0.9720,0.7472,0.7351\} \times c\, R_\infty$, and $d_i=\{0.9036,0.8227,0.8023\}
\times e\, a_0$, respectively, where $m_p$ is the proton mass, $e$ the electron charge,
$a_0$ the Bohr radius, $c$ the speed of light, and $R_\infty$ the Rydberg constant.

In analogy with previous studies of Bose-Hubbard Hamiltonians
\cite{JakBruCir98,GorSanLew02,DemZho02}, it is safe to assume that for a deep enough lattice
and for strictly positive scattering lengths, the ground state of the two-species
Bose-Hubbard model having $N$ atoms per species and $N$ lattice sites will be an insulator
state, i.e. the atoms will be localized in individual wells. The minimum energy arrangement
of the two atomic species then depends on the relative strengths of inter- and intra-species
collision parameters. In principle, there is a broad variety of possible two-species Mott
insulator phases. If, e.g., one species has an extremely weak self-interaction relative to
the other interaction parameters, then it may be energetically favorable for all atoms of
the first species to occupy a single lattice site with the second species uniformly
distributed over the remaining sites. We do not consider such extreme situations and instead
focus on the more likely regime where the ground state contains exactly two atoms per site.
This regime has two phases: a {\it mixed phase} where one atom per species occupies each
site, and a {\it separated phase} where each site contains two atoms of the same species. We
note that the separated phase is unique in that the ground state could be any quantum
superposition of different patterns for distributing the two species.

The collisional energy per lattice of the MI state is given by
$E_{s,m}=\frac{1}{N}\langle\psi_{s,m}|\hat{H}|\psi_{s,m}\rangle\mid_{\{\beta_{ij}\}=0}$,
where $N$ is the number of lattice sites. For the case of two lattice sites, the quantum
state of the separated MI phase is
\begin{equation}
\label{psis}
    |\psi_s\rangle=\frac{1}{2}\hat{c}_{11}^\dag\hat{c}_{11}^\dag\hat{c}_{22}^\dag\hat{c}_{22}^\dag
    |0\rangle,
\end{equation} whereas the ground state of the mixed MI phase is
\begin{equation}
\label{psim}
    |\psi_m\rangle=\hat{c}_{11}^\dag\hat{c}_{12}^\dag\hat{c}_{21}^\dag\hat{c}_{22}^\dag|0\rangle.
\end{equation}
The energy per lattice site calculated from this two-site state will remain valid as the
lattice is scaled up to any even number of sites. The energy per
lattice site for the separated-phase is  $E_s=\frac{1}{2}(g_{11}+g_{22})$, whereas the energy of a pair of
mixed-phase per lattice site is $E_m=g_{12}$. The condition for a mixed-species ground state
is
\begin{figure}
\begin{center}
\centerline{\epsfxsize=3.35in\epsfbox{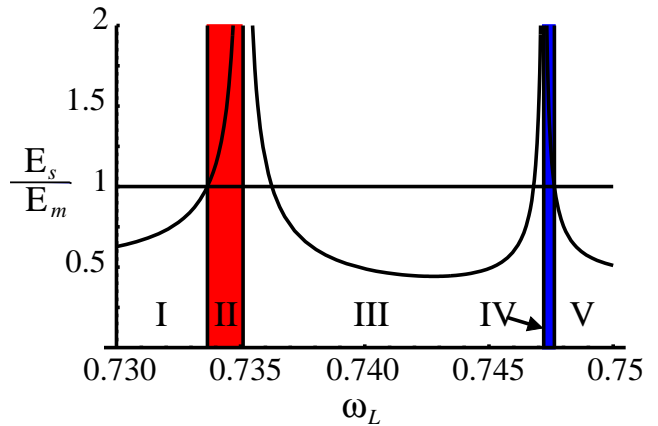}}
\end{center}
\caption{The ratio of the energies per particle of the separated and mixed phases of a
$^{87}$Rb-$^{41}$K two-species Mott insulator as a function of the lattice frequency
$\omega_L$ taken in units of $cR_\infty$.} \label{fig3}
\end{figure}
\noindent therefore $E_s/E_m=(g_{11}+g_{22})/2g_{12}>1$, which can be determined with the
help of Eqs. (\ref{lambdaimu}) and (\ref{gij}). We see that this condition differs
significantly from the free space condition for miscibility \cite{Tim98}.

In Figure \ref{fig3}, we plot this ratio as a function of the lattice frequency
$\omega_L$for the case of $^{87}$Rb and $^{41}$K. We use the scattering length data
$a_{Rb-Rb}=99\, a_0$, $a_{K-K}=60\, a_0$ and $a_{Rb-K}=163 a_0$, taken from the recent
report on a two-species superfluid \cite{ModModRib02}. In figure \ref{fig3}, the frequency
domain is divided into five regions. In region I, both species see a red detuned lattice and
the ground state is a separated dual MI. In region II, which spans from $\omega_L=0.733663$
to the $^{87}$Rb resonance, both species also see a red detuned lattice but the increase in
the rubidium density results in the ground state being the desired mixed phase. This is
because the Rb-Rb mean-field energy is proportional to $\rho_{Rb}^2$, hence the energy cost
of having two $^{87}$Rb atoms can greatly exceed the energy cost of a mixed site despite the
much larger $a_{Rb-K}$. Region III lies between the Rb and K resonances, and hence the two
species see different potential minima. Here the ground state will be a dual MI, but with
the Rb atoms localized at the lattice intensity minima and the K atoms localized at the
intensity maxima. In regions IV, which spans from the K resonance frequency to $\omega_L
=0.747631$ and V, both species see a blue detuned lattice, with region IV corresponding to a
mixed phase and region V to a separated phase.

The mixed phase occurs only in the vicinity of an atomic resonance; thus, we must make sure
that spontaneous emission rates are sufficiently small and remain negligible for
experimentally realizable time scales. By averaging the spontaneous emission rate,
proportional to the light intensity, over the wavefunction (\ref{psii}), we find with the
help of the integral $\int dz\,\left\{
{{\cos^2(kz)}\atop{\sin^2(kz)}}\right\}e^{-z^2}=\frac{\sqrt{\pi}}{2}\left(1\pm
e^{-k^2}\right)$ and  Eqs. (\ref{Vir}) through (\ref{lambdaimu}) that the spontaneous
emission rate of an atom in the ground state of a well of a spherically symmetric lattice
with depth $\alpha \hbar\omega_R$, $\omega_R=\hbar\omega_L^2/(m_ic^2)$ being the atomic
recoil frequency, is given in the Gaussian approximation by
\begin{equation}
    \gamma_i=\frac{3}{2}\alpha\omega_R\frac{\Gamma_i}{|\omega_i-\omega_L|}
    \left(1\pm e^{-1/\sqrt{\alpha}}\right),
\label{gammai}
\end{equation}
where $\Gamma_i$ is the natural linewidth of species $i$ and the $+$ and $-$ signs give the
result for a red-detuned and a blue-detuned lattice, respectively.   For $^{87}$Rb, with
$\Gamma_{Rb}=5$MHz, we find that the average decay rate just to the right of the I-II
boundary is $\gamma_{Rb}\approx \alpha(1+e^{-1/\sqrt{\alpha}})\times 7\times 10^{-2}$Hz,
which for $\alpha=20$ gives a lifetime per atom of $0.38$s. Similarly, on the blue-detuned
side, we find the average decay rate for $^{41}$K, with $\Gamma_K=6$MHz, just to the left of
the IV-V boundary is given by $\gamma_K\approx \alpha(1-e^{-1/\sqrt{\alpha}})\times 7\times
10^{-1}$Hz. For $\alpha=20$ this also gives a lifetime per K atom of $0.35$s, roughly the
same as for Rb. The time scale for adiabatic passage from a SF to a MI state is typically on
the order of $10$ms, so working in either the red (II) or blue (IV) detuned regions should
be feasible without significant spontaneous heating. We note that once the mixed dual-MI
state is reached, the lattice depth could be ramped to a large enough value to suppress
tunneling, after which the lattice frequency could be moved much farther away from the
atomic resonance while preserving the mixed dual-MI state as a metastable state. With this
technique the mixed phase lifetime could be increased significantly to facilitate further
experimentation.

In conclusion, we have calculated the ground state configuration of a two-species optical
lattice for the case where the number of atoms in each species equals the number of lattice
sites and shown that varying the lattice frequency leads to a variety of MI phases. In
making the transition to a dual-MI state from a dual-SF state, it is likely that each
species will undergo the superfluid-insulator transition at a different lattice intensity.
While there are no obvious difficulties
with reaching the mixed dual-MI state in this manner, dynamical simulation of the
Bose-Hubbard model with the exact Wannier states should be performed to ensure that the
system can indeed be driven into the true ground state, a task we plan to carry out in
a future work.

This work was supported by the National Science Foundation through a grant for the Institute
for Theoretical Atomic and Molecular Physics at Harvard University and Smithsonian
Astrophysical Observatory.

Note: During the preparation of this manuscript we became aware of a related manuscript by
Damski {\it et al} \cite{DamSanTie02} which reaches similar conclusions but concentrates
on different aspects of the proposed technique.

\end{document}